\documentclass[aip,jcp,showpacs,showkeys,reprint,floatfix,superscriptaddress]{revtex4-1}

\usepackage{amssymb}
\usepackage{amsmath}
\usepackage{graphicx}
\usepackage{amsbsy}
\usepackage{color}
\usepackage{bm}

\newcommand{\bq}{\begin{eqnarray}}
\newcommand{\eq}{\end{eqnarray}}
\newcommand{\bqn}{\begin{eqnarray*}}
\newcommand{\eqn}{\end{eqnarray*}}

\begin{document}
\title{Quantum Gibbs ensemble Monte Carlo}

\author{Riccardo Fantoni}
\email{rfantoni@ts.infn.it}
\affiliation{Dipartimento di Scienze Molecolari e Nanosistemi,
  Universit\`a Ca' Foscari Venezia, Calle Larga S. Marta DD2137,
  I-30123 Venezia, Italy} 

\author{Saverio Moroni}
\email{moroni@democritos.it}
\affiliation{DEMOCRITOS National Simulation Center, 
Istituto Officina dei Materiali del CNR
and SISSA Scuola Internazionale Superiore di Studi Avanzati, 
Via Bonomea 265, I-34136 Trieste, Italy} 

\date{\today}


\begin{abstract}
We present a path integral Monte Carlo method which is the full 
quantum analogue of the Gibbs ensemble Monte Carlo
method of Panagiotopoulos to study the gas-liquid coexistence line 
of a classical fluid. Unlike previous extensions of Gibbs ensemble
Monte Carlo to include quantum effects, our scheme is viable even 
for systems with strong quantum delocalization in the degenerate 
regime of temperature. This is demonstrated by an illustrative 
application to the gas-superfluid transition of $\mbox{}^4$He 
in two dimensions. 
\end{abstract}

\maketitle
\section{Introduction}
\label{sec:introduction}

Monte Carlo (MC) simulations \cite{Kalos-Whitlock} in the Gibbs
ensemble (GEMC) of Panagiotopoulos  
\cite{Panagiotopoulos87,*Smit89a,*Smit89b}
have now been extensively used for several years to study first order
phase transitions in classical fluids. 
According to the GEMC method, the simulation is performed in two boxes each of
which contains one of two coexisting phases. Equilibration in each phase is
guaranteed by moving particles within the respective box. Equality of
pressures is satisfied in a statistical sense by expanding the volume 
of one of the boxes and contracting the volume of the other. Chemical
potentials are equalized by transferring particles from one
box to the other. This procedure avoids either the laborious search for
matching free energies calculated separately for each phase, or the
simulation of a system large enough to contain both phases and their 
interface.

Notwithstanding the isomorphism between quantum particles and classical 
ring polymers underlying the path integral formulation of quantum 
statistical physics \cite{Feynman1948}, and the recognition that path 
integral Monte Carlo (PIMC) is a tremendously useful numerical 
tool\cite{Ceperley1995} to extract unbiased statistical properties 
of quantum systems, the development of Monte Carlo methods for 
quantum systems is more 
complex, and correspondingly less complete, than for classical ones. 
Putting aside the well known sign problem for fermions \cite{Ceperley1991} 
an important aspect is the development of methods able to simulate a given 
quantum system in different statistical ensembles.    

Recently a new approach to continuous space 
PIMC simulation was devised\cite{Boninsegni2006a,*Boninsegni2006b} 
which makes use of the ``Worm Algorithm'' (WA) previously employed to
study lattice models \cite{Prokofev1998a,*Prokofev1998b}. The WA is
formulated in an enlarged configuration space, which features the
possible presence of an open world-line, the worm. It can 
simulate a system either in the grand canonical or the
canonical ensemble, and it enjoys a favorable scaling of the computational
cost with the system size for the calculation of properties 
related to the formation of long permutation cycles\cite{Pollock1987}, 
such as the superfluid fraction or the one-body density matrix.

It is the purpose of the present work to exploit the WA
\cite{Boninsegni2006b} to obtain an algorithm that is the full
quantum analogue of the GEMC and thus can be used to study 
the gas-liquid phase transition of any (bosonic) quantum
fluid \cite{Young1980}. Several quantum generalizations of GEMC 
have appeared. However, some of them only consider 
particles which have internal quantum states but are otherwise 
classical\cite{Schneider1995,*Nielaba1996}; 
others\cite{Wang1997,*Georgescu2013,*Kowalczyk2013} are limited to 
particles isomorph to relatively compact classical polymers
(hence high enough temperature and/or small enough quantumness); 
none of them features the structure of particle exchanges which 
underlies Bose (or Fermi) statistics.
We apply the quantum Gibbs ensemble 
Monte Carlo (QGEMC) method to the liquid-gas coexistence of
two-dimensional $^4$He where strong quantum effects, including
superfluidity, are present. 


\section{Classical Gibbs Ensemble Monte Carlo}
\label{sec:cgemc}
We begin with a brief summary of the Gibbs Ensemble Monte Carlo method,
that we deem useful for the subsequent quantum generalization.
A detailed presentation is given in Ref. \onlinecite{Frenkel-Smit}.

The system comprises a box of volume $\Omega_1$ containing $N_1$ particles
and a box of volume $\Omega_2$ containing $N_2$ particles.
The temperature $T$, the total number of particles 
$N=N_1+N_2$, and the total volume $\Omega=\Omega_1+\Omega_2$ are fixed, and there is no 
interaction between particles enclosed in different boxes.
Starting from the partition function for the Gibbs ensemble
\begin{equation}
Z_G(N,\Omega,T)=\frac{1}{\Omega}\sum_{N_1=0}^N\int d\Omega_1
Z(N_1,\Omega_1,T)Z(N_2,\Omega_2,T),
\label{Z_gibbs}
\end{equation}
where $Z$ is the canonical partition function,
the probability density for the coordinates 
$R=\{{\bf r}_1,\ldots,{\bf r}_N\}$ of the particles,
the number $N_1$ and the volume $\Omega_1$ can be cast in the form
\begin{equation}
P_{N,\Omega,T}(R,N_1,\Omega_1)\propto \frac{\Omega_1^{N_1+1}\Omega_2^{N_2+1}}{N_1!N_2!}
e^{-\beta V(R)}.
\label{p_classical}
\end{equation}
Here $\beta=1/k_BT$ and the potential energy in the Boltzmann weight,
assuming a central pair potential $v(r)$, is
\begin{equation}
V(R)=\sum_{i=1}^{N_1-1}\sum_{j=i+1}^{N_1}v(r_{ij})
    +\sum_{i=N_1+1}^{N-1}\sum_{j=i+1}^{N}v(r_{ij}).
\label{u_classical}
\end{equation}
The Monte Carlo simulation proceeds via three kinds of moves:

(1) Displace the position ${\bf r}_i$ of a randomly selected particle
within its own box; this is done as in standard canonical ensemble 
simulations.

(2) Change the volumes; this is done by uniformly sampling a displacement 
of the quantity ${\rm ln(\Omega_1/\Omega_2)}$, with $\Omega$ kept fixed.

(3) Exchange particles; this is done by transferring a randomly
chosen particle to a random position in the other box.

The acceptance probabilities are obtained
imposing detailed balance\cite{Frenkel-Smit}.
After equilibration, provided $N/\Omega$ is within the coexistence 
region at the temperature $T$, each of the two boxes will contain 
one of the coexisting phases.

\section{Quantum Gibbs Ensemble Monte Carlo}
\label{sec:qgemc}
The QGEMC is based on the Path Integral Monte Carlo method in the Worm
Algorithm implementation. We refer to the literature 
\cite{Ceperley1995,Boninsegni2006b} for a full account of these techniques, 
giving here only a brief discussion of some aspects relevant
to the quantum generalization of the classical GEMC.

\subsection{Path Integral Monte Carlo}
\label{subsec:pimc}
We consider an assembly of $N$ identical particles
obeying Bose statistics. In the position representation, 
the canonical partition function is 
\begin{equation}
Z= \frac{1}{N!} \sum_{\cal P} \int dR\rho(R,{\cal P}R;\beta),
\label{canonical_Z}
\end{equation}
where $\rho(R,R';\beta)=\langle R|e^{-\beta H}|R'\rangle$ is the thermal
density matrix for distinguishable particles, and the sum over
the permutations ${\cal P}$ accounts for Bose symmetry.
The density matrix can be expressed in a form amenable 
to Monte Carlo simulation in terms of discretized path 
integrals:
\begin{equation}
\rho(R,R';\beta)\simeq\int dR_1\ldots dR_{K-1} \prod_{j=1}^K {\tilde\rho}
(R_{j-1},R_j;\epsilon),
\label{rho_beta}
\end{equation}
with $R_0=R$, $R_K=R'$, and $\{R_1,\ldots,R_{K-1}\}$ a sequence (path) of 
intermediate configurations. An adjacent pair $\{R_{j-1},R_j\}$ is called a 
{\it link}. In Eq. (\ref{rho_beta}) the factors ${\tilde \rho}$
have an argument $\epsilon=\beta/K$ which corresponds to a
temperature $K$ times higher than $T$, and for high temperature the unknown
many-body density matrix can be accurately approximated by an explicit 
expression of the general form 
\begin{equation}
{\tilde\rho}(R,R';\epsilon)=\rho_F(R,R';\epsilon)e^{-U(R,R';\epsilon)},
\label{rho_tilde}
\end{equation}
where 
\begin{equation}
\rho_F(R,R';\epsilon)=
(4\pi\lambda\epsilon)^{-dN/2}
\prod_{i=1}^N
e^{-({\bf r}_i-{\bf r}'_i)^2/4\lambda\epsilon}
\label{rho_free}
\end{equation}
is the density matrix for $N$ non-interacting particles in $d$ spatial
dimensions, and the function $U$ takes into account the effect of correlations.
In the limit $\epsilon\to 0$, ${\tilde\rho(R,R',\epsilon)}$ approaches 
$\rho(R,R',\epsilon)$ and the approximate equality (\ref{rho_beta}) 
becomes exact.

For each particle, Eq. (\ref{rho_beta}) defines a trajectory, or world
line (WL),
$\left\{{\bf r}_{i;0},{\bf r}_{i;1},\ldots,{\bf r}_{i;K} \right\}$,
where the {\it bead} ${\bf r}_{i;j}$ is the position of the $i$th particle
at the $j$th ``time'' discretization 
index. In the calculation of thermal averages, 
$\langle A\rangle=Tr\rho A/Z$, the presence of the traces 
and the Bose symmetry of Eq. (\ref {canonical_Z}) require 
periodic boundary conditions in time, 
${\bf r}_{i;K}={\bf r}_{{\cal P}i;0}$: the trajectory of a 
particle ends in the initial position of either the same or 
another particle, according to the permutation cycles contained
in the permutation ${\cal P}$. All
the interlinked trajectories of a permutation cycle of $k$ particles
form a single WL with $kK$ steps, so that all WLs are closed. 
The WL of a single particle
has a spatial extent limited by the thermal 
wawelength, while the WL of a long permutation cycle 
can span the whole system.

The simulation proceeds by sampling a density probability 
proportional to the integrand of Eq. (\ref{rho_beta}). Specific
techniques are devised to update not only the particle positions
along the WLs, but also the permutations.

The WLs can be mapped onto classical ring polymers, with peculiar
interactions defined through Eq (\ref{rho_beta}) by viewing the
integrand as a Boltzmann weight. Thus it seems possible to apply
the GEMC method to the quantum system as well.
However, an issue arises with the exchange move: a quantum
particle corresponds, in the classical mapping, to a whole polymer,
and the acceptance rate for transferring a polymer to the other box
can be expected to be low, particularly at low temperature when 
the thermal wavelength increases and the spatial extension of 
the polymers grows. The problem is further compound by the 
presence of interlinked trajectories belonging to a permutation cycle.  
This is why quantum applications of GEMC have been limited to 
relatively high temperature and/or relatively low 
quantumness\cite{Wang1997,*Georgescu2013,*Kowalczyk2013}.
We will show how to overcome these difficulties using the WA.

\subsection{Worm Algorithm}
\label{subsec:wa}
The WA enlarges the configuration space:
along with the closed WLs of Section \ref{subsec:pimc},
there are configurations with an open WL in which one 
particle is created in ${\bf r}_{\cal M}$ at time
$j_{\cal M}\epsilon$ and destroyed in ${\bf r}_{\cal I}$ 
at a later time $j_{\cal I}\epsilon$. The difference
$j_{\cal I}-j_{\cal M}$ is intended modulo $K$,
and the open WL can belong to a permutation cycle involving other particles. 
The points ${\bf r}_{\cal I}$ and ${\bf r}_{\cal M}$ are called
Ira and Masha, respectively, and the WL connecting them is called the 
{\it worm}.
Configurations with only closed WLs belong to the ``Z sector''
and contribute to the partition function. Configurations with
a worm belong to the ``G sector'' and contribute to the
one-body Green function 
$g({\bf r}_{\cal M},{\bf r}_{\cal I};(j_{\cal I}-j_{\cal M})\epsilon)/Z$.
All physical properties, with the exception of the Green function, are
calculated only on configurations of the Z sector.
The full set of configurations corresponds to the extended partition
function 
\begin{equation}
Z_W=Z+Z'
\label{extended_Z}
\end{equation}
where $Z$ can be either the canonical or the grand partition function,
\begin{equation}
Z'=C\sum_{j_{\cal I},j_{\cal M}}\int d{\bf r}_{\cal I} d{\bf r}_{\cal M}
g({\bf r}_{\cal M},{\bf r}_{\cal I};(j_{\cal I}-j_{\cal M})\epsilon),
\label{Z_g}
\end{equation}
and the arbitrary parameter $C$ defines the relative weight of the Z 
and G sectors. The discretized path integral 
expression of Eq. (\ref{Z_g}) is obtained
in close analogy with Section \ref{subsec:pimc} in terms of 
${\tilde \rho}(R_{j-1},R_j;\epsilon)$.

The simulation proceeds via a set of local moves --the complementary pairs 
{\it Open} and {\it Close},
{\it Insert} and {\it Remove}, {\it Advance} and {\it Recede}, and the
self-complementary {\it Swap}--
which guarantee ergodic sampling ot the enlarged configuration space
by switching between the Z and the G sectors and displacing the
coordinates of the particles \cite{Boninsegni2006b}.

The usefulness of the WA for the implementation of the QGEMC can be
appreciated by considering the process of adding a particle to the system
(we assume here that $Z$ is the grand partition function):
starting from the Z sector, a worm may be {\it inserted};
once in the G sector, the worm may 
{\it advance}, possibly {\it swap} with existing closed WLs, and
eventually get {\it closed}, thus switching back to the Z sector with 
one more particle. Each single move is a local update that 
involves only a limited number of time steps, so that the acceptance rate 
can be high even in a dense system.

\subsection{Gibbs Ensemble}
\label{subsec:gibbs}
We consider $N_1$ particles in a volume $\Omega_1$ and $N_2$ particles 
in a volume $\Omega_2$, with $\Omega$, $N$, and $T$ fixed (see 
Section \ref{sec:cgemc}). The configurations of the system in the
Gibbs ensemble are distributed
according to the partition function $Z_G$ of Eq. (\ref{Z_gibbs}), with 
each of the canonical partition functions $Z$ of the two subsystems expressed 
as discretized path integrals with closed WLs, as in Section 
\ref{subsec:pimc}. These configurations define the Z sector.

\begin{figure}[h]
\begin{center}
\includegraphics[width=6cm,angle=270]{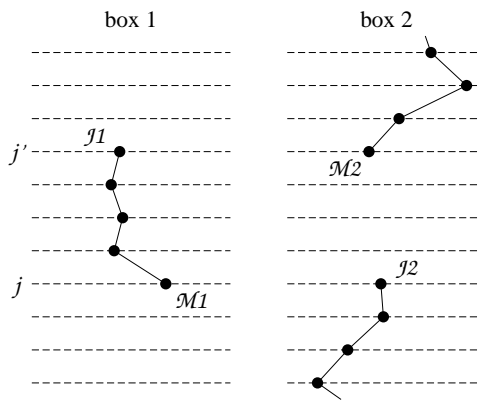}
\end{center}  
\caption{
Schematic illustration of open WLs in the G sector.
}
\label{fig:g_sector}
\end{figure}

Following the strategy of the WA we enlarge the configuration 
space allowing for open WLs, while strictly enforcing the constraint 
of fixed $N$: whenever there is a worm in box 1, 
with Masha at ${\bf r}_{{\cal M}1;j}$ and Ira at 
${\bf r}_{{\cal I}1;j'}$,
there is a worm in box 2 as well, with
Masha at ${\bf r}_{{\cal M}2;j'}$ and Ira at 
${\bf r}_{{\cal I}2;j}$, as schematically illustrated in Fig. 
\ref{fig:g_sector}. These configurations define the G sector.
In the G sector the number of particles in box $\alpha$ ($\alpha=1,2$)
varies between $N_\alpha$ and $N_\alpha-1$, with $N_1+N_2=N+1$,
and the total number of particles within each link is $N$.

The extended partition function is $Z_W=Z_G+Z'$, where
\begin{equation}
Z'=\frac{1}{\Omega}\sum_{N_1}\int d\Omega_1 C \sideset{}{'}\sum_{j,j'=1}^KF_1(j,j')F_2(j',j).
\label{Z_prime}
\end{equation}
The primed summation excludes the terms with $j=j'$ to make sure
there is a worm per box in the G sector.
and the function $F_\alpha$ --the integral of
Eq. (\ref{Z_g}) for box $\alpha$-- is 
expressed in terms of density matrices as 
\begin{widetext}
\begin{equation}
F_\alpha(j,j')=\frac{1}{N_\alpha!}\sum_{{\cal P}_\alpha}\int 
\rho\left(\{R_\alpha,{\bf r}_{{\cal M}\alpha}\}
,{\cal P}_\alpha\{R_\alpha',{\bf r}_{{\cal I}\alpha}\};\tau_{j,j'}\right)
\rho\left(R'_\alpha,R_\alpha;\tau_{j',j}\right)
dR_\alpha dR_\alpha' d{\bf r}_{{\cal M}\alpha} d{\bf r}_{{\cal I}\alpha}.
\label{lunga}
\end{equation}
\end{widetext}
Here the pair $\{R_\alpha,{\bf r}_{{\cal M}\alpha}\}$ indicates
the coordinates of Masha and of all the other particles of
box $\alpha$ at time index $j$ (the first argument of $F_\alpha$) and 
$\{R'_\alpha,{\bf r}_{{\cal I}\alpha}\}$ the coordinates of Ira and
of the other particles at $j'$.
The argument $\tau_{j,j'}$ of the
density matrices $\rho$ is the positive interval from $j\epsilon$ 
to $j'\epsilon$ --possibly wrapping around the periodic boundary condition, 
i.e. $\tau_{j,j'}=\left [(j'-j+K)~{\rm mod}~K\right ]\epsilon$.
Finally, the density matrices are expanded in discretized path integrals
using the high temperature approximation ${\tilde\rho}$ as in 
Section \ref{subsec:pimc}.

The probability density for all the coordinates $X$ in the system, the
number $N_1$ and the volume $\Omega_1$ is\cite{note2}
\begin{equation}
P_{N,\Omega,T}(X)\propto C^{\delta_G}\prod_{j=1}^K {\tilde\rho}(X_{j-1},X_j;\epsilon),
\label{p}
\end{equation}
where $\delta_G$ is 1(0) in the G(Z) sector, $X_j$ indicates the
positions of all the particles in either box at time $j\epsilon$, 
and the dependence on $N_1$ and $\Omega_1$, as well as all possible 
permutations of particle labels, are implicitly contained in the
configuration $X$. 
No 
sums over permutations appear in $P$ because 
the symmetrizations of Eqs. (\ref{lunga}) or (\ref{canonical_Z})
are carried out concurrently with the Monte Carlo integration
over the coordinates, through updates of the permutation cycles.

We next describe a set of moves which sample the configuration 
space with probability density $P_{N,\Omega,T}(X)$.
They are the standard moves of PIMC and the WA, in some cases combined 
in pairs to preserve the two-worm structure of the G sector
illustrated in Fig. \ref{fig:g_sector}, and the volume change
move specific of the GEMC method; the particles exchange move of GEMC
builds spontaneously through a sequence of WA moves. The acceptance 
probabilities are obtained by enforcing detailed balance according
to the generalized Metropolis algorithm \cite{Frenkel-Smit}
(if the current configuration is in a sector where the proposed 
move is not applicable, the move is rejected immediately).

\begin{figure}[h]
\begin{center}
\includegraphics[width=6cm,angle=270]{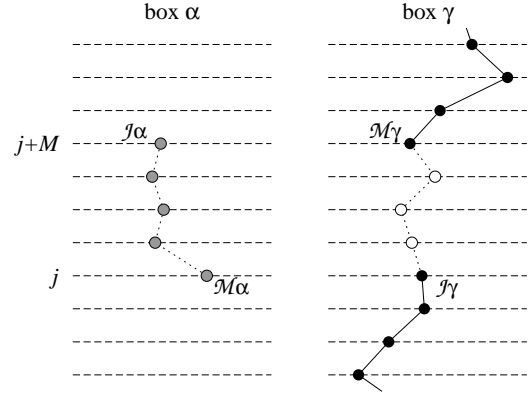}
\end{center}  
\caption{
Schematic illustration of the open-insert move. Two worms are created by 
removing the white beads and inserting the grey beads.
}
\label{fig:insert_open}
\end{figure}

{\it (1a) Open-insert}.
This move, schematically illustrated in Fig. \ref{fig:insert_open},
is applicable only in the Z sector. It switches from the Z to the 
G sector by opening an existing closed WL in one box and inserting 
a new open WL in the other box.
A particle is picked randomly, and the links of its WL from $j$ to $j+M$
are removed. The time index $j$ is 
uniformly sampled in $[1,K]$, and the number of removed links $M$
is uniformly sampled in $[1,{\bar M}]$, where ${\bar M}<K$
is a parameter of the simulation which controls the size of the move.
Let $\alpha$ be the label of the other box.
The initial bead ${\bf r}_{{\cal M}\alpha;j}$ of the new WL is placed
at a position ${\bf r}_0$ randomly sampled in $\Omega_\alpha$, and 
$M$ further beads are sampled from 
$\prod_{\nu=1}^M\rho_0({\bf r}_{\nu-1},{\bf r}_{\nu};\epsilon)$, where
\begin{equation}
\rho_0({\bf r},{\bf r}';\epsilon)=(4\pi\lambda\epsilon)^{-d/2}
e^{-({\bf r}-{\bf r}')^2/4\lambda\epsilon}
\end{equation}
is the one-particle free propagator. 
The acceptance probability is 
$p_{\rm op-in}={\rm min}\left\{1,e^{\Delta_U}\pi_{\rm op-in}\right\}$,
where
\begin{equation}
\pi_{\rm op-in}=\frac{C{\bar M}K\Omega_\alpha N}
         {2\rho_0({\bf r}_{{\cal I}\gamma},{\bf r}_{{\cal M}\gamma};M\epsilon)}
\label{pi}
\end{equation}
and $\Delta_U=\sum_{\nu=1}^M \left[ U(X_{\nu-1},X_{\nu};\epsilon)
                            -U(X^*_{\nu-1},X^*_{\nu};\epsilon)\right]$ is
the change of the interacting part of the action $U$ between the initial 
configuration $X$ and the proposed configuration $X^*$.

{\it (1b) Close-remove} is the complementary move of {\it open-insert}.
A box --say $\gamma$-- is selected at random. If 
$M=\tau_{{\cal I}\gamma,{\cal M}\gamma}/\epsilon >{\bar M}$, 
the move is rejected. Otherwise,
a WL of $M$ links connecting ${\bf r}_0={\bf r_{{\cal I}\gamma}}$ to 
${\bf r}_M={\bf r_{{\cal M}\gamma}}$ is sampled from 
$\prod_{\nu=1}^M\rho_0({\bf r}_{\nu-1},{\bf r}_{\nu};\epsilon)$.
If the open WL in the other box contains more than ${\bar M}$ links
the move is rejected, otherwise the worm is removed. The acceptance 
probability is 
$p_{\rm cl-rm}={\rm min}\left\{1,e^{\Delta_U}/\pi_{\rm op-in}\right\}$.

{\it (2) Advance-recede}. This move is self-complementary, as are all 
the remaining moves. It applies only to the G sector, and we refer to
Fig. \ref{fig:g_sector} for a representation of the initial configuration. 
A box --say $\gamma$-- is selected at random.
An integer $M$ is uniformly sampled in 
$\left[ 1,{\bar M}\right]$ and a time direction is selected at random. 
If the time direction is positive, 
a new portion of WL sampled from a product of $M$ free-particle propagators
is added in box $\gamma$ starting
from ${\bf r_{{\cal I}\gamma}}$,
and a corresponding, $M$-link portion of the open WL 
existing in box $\alpha$ is removed, starting from ${\bf r_{{\cal M}\alpha}}$.
If the time direction is negative, the new portion of WL is added
in box $\gamma$ starting from ${\bf r_{{\cal M}\gamma}}$ and going 
backwards in time, and the WL in box $\alpha$ is shortened
starting from ${\bf r_{{\cal I}\alpha}}$. The move is rejected if
$M\geq\tau_{{\cal I}\gamma,{\cal M}\gamma}/\epsilon$
(this restriction could be avoided using more elaborate combinations
of the WA moves). The acceptance probability of {\it advance-recede} is
$p_{\rm ad-re}={\rm min}\left\{1,e^{\Delta_U}\right\}$.

{\it (3) Swap}. This move applies only to the G sector. A box is selected
at random, and within the chosen box the move proceeds in the same way as
in the WA \cite{Boninsegni2006b}.

{\it (4) Volume change}. We choose to apply this move only to 
configurations of the Z sector.
For the classical GEMC update of the volumes, it proves convenient to make
the dependence on $N_1$ and $\Omega_1$ explicit. This is achieved
\cite{Frenkel-Smit} by rescaling all lengths in box $\alpha$
by $\Omega_\alpha^{-1/d}$ and formally performing the Monte Carlo
integration over the rescaled coordinates $\Xi(X)$. Furthermore,
the move is usually implemented \cite{Frenkel-Smit} by changing
the quantity ${\rm ln(\Omega_1/\Omega_2)}$, rather than $\Omega_1$
itself, by an amount uniformly sampled in $[-\Delta_\Omega,\Delta_\Omega]$ 
with $\Delta_\Omega$ a parameter which controls the size of the move. A
factor $\Omega_1^{N_1}\Omega_2^{N_2}$ appears in 
$P_{N,\Omega,T}$ as a result of rescaling the coordinates, and another factor 
$\Omega_1\Omega_2$ as a result of updating the logarithm of the volume
(cfr. Eq. (\ref{p_classical})). In the quantum case we adopt
the same changes of variables. Since
each particle is mapped onto $K$ beads, each of which gets rescaled 
coordinates, the probability density is
\begin{eqnarray}
P_{N,\Omega,T}(\Xi,N_1,\Omega_1)&\propto &\Omega_1^{KN_1+1}\Omega_2^{KN_2+1}\nonumber \\
&\times&\prod_j{\tilde\rho}(X_{j-1}(\Xi),X_j(\Xi);\epsilon).
\label{p_vol}
\end{eqnarray}
The acceptance probability for a move from $\Omega_1$ to $\Omega_1^*$ is
\begin{equation}
\label{acc_vol}
p_{\rm vol}={\rm min}
\left\{ 1,\left(\frac{\Omega_1^*}{\Omega_1}\right)^{KN_1+1}
\left(\frac{\Omega_2^*}{\Omega_2}\right)^{KN_2+1} 
e^{\Delta_S}\right\},
\end{equation}
where $\Delta_S$ is the change of the {\it full} action between 
the initial configuration $X$ and the proposed configuration 
$X^*$:
\begin{equation}
\Delta_S=-\sum_{\nu=1}^K
\ln \left[\tilde\rho(X_{\nu-1},X_{\nu};\epsilon)
/\tilde\rho(X^*_{\nu-1},X^*_{\nu};\epsilon)\right].
\label{delta_s}
\end{equation}
The proposed configuration is
$X^*=(\Omega_\alpha^*/\Omega_\alpha)^{1/d}X$, with $\alpha=1$ or $2$
as appropriate to the particle index of each component of $X$. Hence, both 
the equal-time interparticle distances, $|{\bf r}_{i;j}-{\bf r}_{k;j}|$,
and the single-particle displacements along the WL, 
$|{\bf r}_{i;j-1}-{\bf r}_{i;j}|$, are modified upon volume changes.
This prescription departs from that recommended for classical
systems of composite particles \cite{Frenkel-Smit}, where only 
the center of mass follows the variation of the volume while the 
internal structure remains unchanged (in the quantum analogue, 
only the centroid ot each ring polymer would change while the size 
and shape of the polymers would stay fixed \cite{Wang1997}).
The reason for the prescription chosen here is that for 
polymers interlinked through permutation cycles the equal-time
interparticle distance and the single-particle paths are not
independent.

In our implementation we also include moves which {\it wiggle} an existing
portion of a WL, or {\it displace} the whole WL of a particle. These moves
are standard in PIMC \cite{Ceperley1995} and since they are not strictly 
needed for the QGEMC we do not describe them here.

\section{Two-dimensional $\mbox{}^4$He}
\label{sec:4He}
The phase diagram of $^4$He in two dimensions has been studied 
in Ref.~\onlinecite{Gordillo1998} by
PIMC simulations of individual phases for many values of
density and temperature. A gas-liquid coexistence 
region is found below 0.87 K. At these temperatures, on account of the
large De Boer parameter of $^4$He, $\Lambda=0.429$ \cite{Young1980}, 
quantum exchange of particles is an important 
effect\cite{Pollock1987,Ceperley1995}:
in the thermodynamic limit the normal-superfluid transition temperature
at saturated vapour pressure is 0.65 K \cite{Boninsegni2006b}, and for
finite systems of a few hundred particles the superfluid fraction
is non zero even for $T=1$~K. Therefore the gas-liquid coexistence of 
two-dimensional $^4$He is a telling test of the QGEMC algorithm for
a degenerate quantum system.

We simulate a two-dimensional system of $N=64$ $^4$He atoms distributed 
between two square boxes with periodic boundary conditions. Within each
box, the atoms interact with the HFDHE2 pair potential \cite{Aziz1979}.
We use the primitive approximation
\begin{equation}
{\tilde \rho}(R,R';\epsilon)=\rho_F(R,R';\epsilon)e^{-\epsilon[V(R)+V(R')]/2}
\label{primitiva}
\end{equation}
to the high temperature density matrix, with $\epsilon=0.002$ K$^{-1}$.
We study the temperature range between 0.125 K and 1 K. For each
temperature, the simulation starts from a configuration with boxes
of equal volume containing 32 atoms each at a density 0.025~\AA$^{-2}$.

\begin{figure}[htbp]
\begin{center}
\includegraphics[width=7cm]{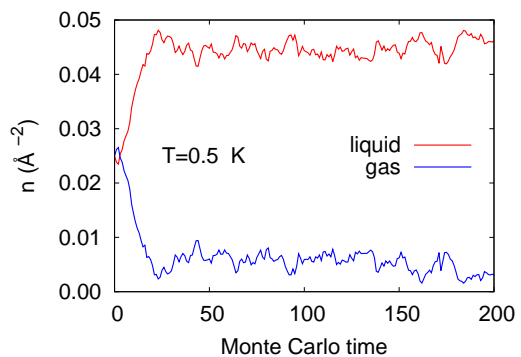}
\end{center}  
\caption{
(color online) Data trace of the densities of the gas 
(blue) and the liquid (red) in the initial stages of 
the simulation for $T=0.5$ K.
}
\label{fig:den}
\end{figure}

\begin{figure}[htbp]
\begin{center}
\includegraphics[width=7cm]{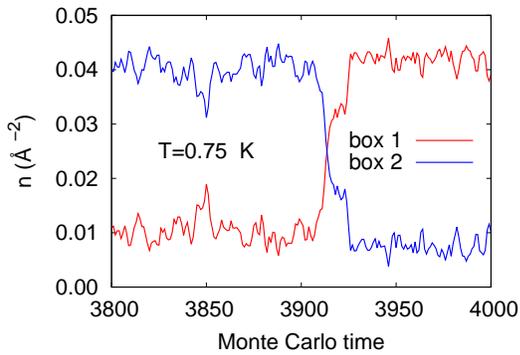}
\end{center}  
\caption{
(color online) A portion of the data trace of the densities 
for the simulation at $T=0.75$ K, showing an exchange of 
identity between the two boxes.
}
\label{fig:den2}
\end{figure}

After equilibration, deep in the subcritical temperature regime
one of the boxes contains a gas of very low density $n_g$, and
the other a superfluid liquid with a density $n_l$ close to 
the equilibrium density of the system at $T=0$ (see Fig. \ref{fig:den}).
For temperatures closer to the critical point, $n_g$ and $n_l$ 
approach each other, and we frequently observe
that the two boxes exchange identity, i.e. the phase of the system 
in each box switches back and forth between gas and liquid 
(see Fig. \ref{fig:den2}). In this case the density
has a bimodal distribution peaked at the values $n_g$ and $n_l$ 
of the coexisting phases.
This bimodal distribution can be obtained in a grand canonical simulation
of a single box, but this requires a fine tuning of the chemical 
potential\cite{Wilding1995}.
For $T=1$~K the two peaks merge into a single gaussian 
centered at the average density 0.025~\AA$^{-2}$.


\begin{figure}[htbp]
\begin{center}
\includegraphics[width=7cm]{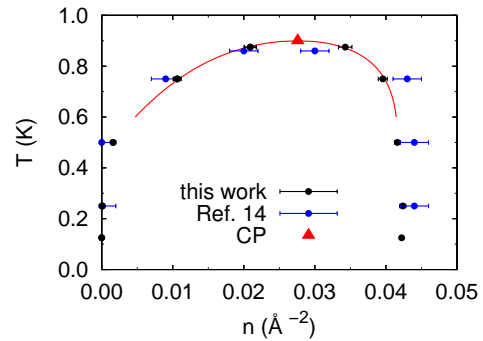}
\end{center}  
\caption{
  (color online) The binodal line of $\mbox{}^4$He in two 
  dimensions. Black points: QGEMC. Blue points:
  Ref.~\onlinecite{Gordillo1998}. Red line: 
  extrapolation of the QGEMC results to the critical point (red triangle).
  }  
\label{fig:pd}
\end{figure}

Our results for the densities $n_g$ and $n_l$ of the coexisting phases
are listed in Table~\ref{tab:2d4he} and displayed in Fig.~\ref{fig:pd}.
They compare favorably with the results of Ref.~\onlinecite{Gordillo1998}.
For each $T$, the latter are
obtained from an integral of the isothermal pressure 
calculated in the canonical ensemble for several values of the density
across the coexistence region; the QGEMC method is simpler because
$n_g$ and $n_l$ are obtained with a single simulation, either
directly or via the analysis of a bimodal distribution.
Each of the present QGEMC calculations took $\sim 300$ CPU hours on a 2GHz 
processor. If needed, the efficiency could be significantly improved 
using a better approximation to the high temperature density 
matrix\cite{Ceperley1995}.

 
%
\begin{table}[htbp]
\caption{
  The densities $n_g$ and $n_l$ of the coexisting gas and liquid phases
  as a function of the temperature $T$.
  } 
\label{tab:2d4he}
{
\begin{ruledtabular}
\begin{tabular}{cll}
$T$~(K) & $n_g$~(\AA$^{-2}$) & $n_l$~(\AA$^{-2}$) \\ 
\hline
\multicolumn{3}{c}{this work} \\ 
\hline
0.125 & $<$10$^{-6}$ & 0.0422(2) \\
0.250 & 0.00009(2) & 0.0424(4) \\
0.500 & 0.0016(2)  & 0.0416(4) \\
0.750 & 0.0106(6)  & 0.0396(7) \\
0.875 & 0.0209(9)  & 0.0343(10) \\
\hline
\multicolumn{3}{c}{Ref. \onlinecite{Gordillo1998}} \\ 
\hline
0.250 & 0.000(2) & 0.044(2) \\
0.500 & 0.000(2) & 0.044(2) \\
0.750 & 0.009(2) & 0.043(2) \\
0.860 & 0.020(2) & 0.030(2) \\
\end{tabular}
\end{ruledtabular}
}
\end{table}

The boundary of the gas-liquid coexistence region is called the binodal line.
It can be extrapolated to the critical point (CP) using the law of 
``rectilinear diameters'' \cite{Guggrnheim1945},
$\rho_l+\rho_g=2\rho_c+a|T-T_c|$, and  the expansion \cite{Fisher1966}
$\rho_l-\rho_g=b|T-T_c|^{\beta_1}(|T-T_c|+c)^{\beta_0-\beta_1}$. Here
$\beta_1=1/2$ and $\beta_0=1/8$, while $\rho_c,~ T_c~, a~, b$ and $c$ are
fitting parameters. We find $\rho_c=0.028$~\AA$^{-2}$ and $T_c=0.90$~K.

%

Finally we list in Table \ref{tab:rhos} the winding number
estimator \cite{Pollock1987} of the superfluid fraction $n_s$
for the two phases as a function of the temperature.
A non zero value for the liquid branch of the binodal over 
the full range of temperatures considered is a clear indication 
of the importance of quantum exchanges (although, as mentioned, 
a finite value of $n_s$ for $T>$ 0.65 K
is a finite size effect \cite{Boninsegni2006b}).

\begin{table}[htbp]
\caption{
  The superfluid fraction $n_s$ in the coexisting gas and fluid phases as
a function of $T$.
  }  
\label{tab:rhos}
{
\begin{ruledtabular}
\begin{tabular}{lll}
$T$~(K) & \multicolumn{2}{c}{$n_s$} \\ 
         & gas & liquid \\ 
\hline
1.000 & \multicolumn{2}{c}{0.032(1)} \\
0.875 & 0.14(3) & 0.36(2) \\
0.750 & 0.06(3) & 0.63(3) \\
0.500 & 0.0014(5)&  0.938(7) \\
0.250 & $<$10$^{-3}$ & 0.963(10) \\
0.125 & 0 & 0.985(16) \\
\end{tabular}
\end{ruledtabular}
}
\end{table}
%

\section{Conclusions}
\label{sec:conclusions}

We have presented the QGEMC method, a full quantum extension of classical 
Gibbs Ensemble Monte Carlo based on the Worm Algorithm. The method is 
demonstrated for the binodal of $^4$He in two dimensions, a physical
property of a strongly quantum system in the degenerate temperature regime. 
Good agreement is found with the results of previous PIMC
simulations in the canonical ensemble. In analogy with applications 
of GEMC to classical fluids\cite{Fantoni2011,Fantoni2013a,Fantoni2013},
the QGEMC method offers a convenient approach
for problems such as gas-liquid coexistence in quantum systems 
and phase equilibria in quantum mixtures.


\begin{acknowledgments}
We are greatful to M. E. Fisher for correspondence and helpful comments.
\end{acknowledgments}
\bibliographystyle{apsrev}

\begin{thebibliography}{28}
\bibitem[{not()}]{note2}
\bibinfo{note}{The link density matrix ${\tilde \rho}(X,X';\epsilon)$ of the
  whole system is the product ${\tilde \rho}(R_\alpha,R'_\alpha;\epsilon)
  {\tilde \rho}(R_\gamma,R'_\gamma;\epsilon)$ over the two boxes; particle
  labels are only permuted within each box.}
\end{thebibliography}

\end{document}